\documentclass[preprint,12pt]{elsarticle}

\usepackage{amssymb}
\usepackage{amsmath}

\usepackage{graphicx}
\usepackage{subfig}
\usepackage{textcomp}
\usepackage{siunitx}
\usepackage{makeidx}

\journal{Chaos, Soliton \& Fractals}

\newcommand{\fig}[1]{Fig. \ref{#1}}

\newcommand{\eq}[1]{Eq. (\ref{#1})}
\newcommand{\pico}[1]{#1}

\begin{document}

\begin{frontmatter}
	
	
	
	\title{Langevin’s model for soliton molecules in ultrafast fiber ring laser cavity: investigating experimentally the interplay between noise and inertia}
	
	
\author[a]{Anastasiia Sheveleva}
\author[a]{Aur\'{e}lien Coillet}
\author[a]{Christophe Finot}
\author[a,b]{Pierre Colman}

\affiliation[a]{organization={Laboratoire Interdisciplinaire Carnot de Bourgogne, UMR 6303 CNRS, Universite de Bourgogne},
	addressline={9 avenue Alain Savary}, 
	city= {Dijon},
	country={FRANCE}}

\affiliation[b]{organization={Corresponding author: pierre.colman@u-bourgogne.fr},
	}
	
\begin{abstract}
	The dynamics of soliton molecules in ultrafast fiber ring laser cavity is strongly influenced by noise. We show how a parsimonious Langevin model can be constructed from experimental data, resulting in a mathematical description that encompasses both the deterministic and stochastic properties of the evolution of the soliton molecules. In particular, we were able to probe the response dynamics of the soliton molecule to an external kick in a sub-critical approach, namely without the need to actually disturb the systems under investigation. Moreover, the noise experienced by the dissipative solitonic system, including its distribution and correlation, can now be also analyzed in details. Our strategy can be applied to any systems where the individual motion of its constitutive particles can be traced; the case of optical solitonic-system laser presented here serving as a proof-of-principle demonstration.
\end{abstract}

\begin{graphicalabstract}
	\includegraphics[width=1\linewidth]{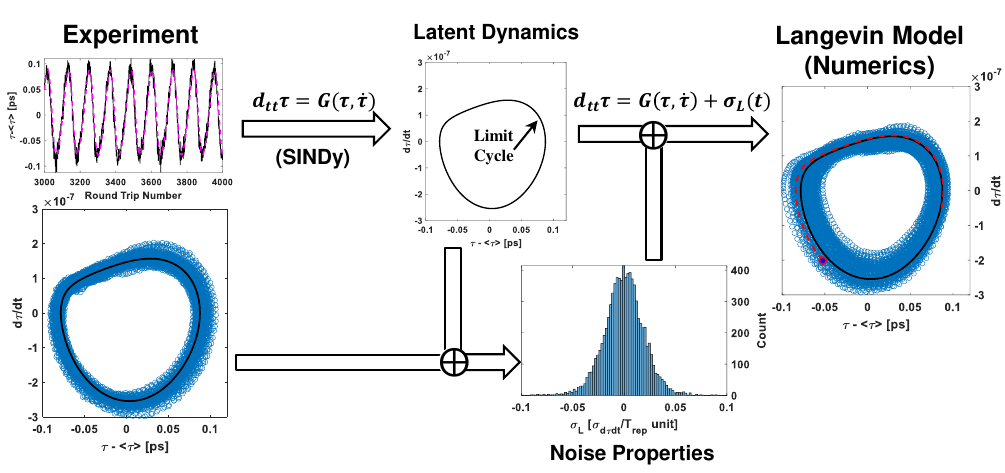}
\end{graphicalabstract}

\begin{highlights}
	\item Parsimonious dynamical equations can be derived from experimental data, despite the presence of noise
	\item Compared to the investigation of an ultra-stable system, random fluctuations actually help extending further the validity range of any numerical model fitting the data. This is in particular a key point when investigating oscillators.
	\item A numerical twin allows to investigate separately the latent dynamics and the noise (fluctuations), while the two cannot be measured independently from each other.
\end{highlights}

	
\end{frontmatter}



\section{Introduction}

\par The subtle balance between gain and loss mechanism, and dispersion and nonlinearity in an ultrafast fiber ring laser cavity can result in the spontaneous generation of very specific invariant pulses known as dissipative solitons \cite{Grelu2012}. A few of these co-propagating pulses can interact closely and form a bound state called the soliton molecule (SM). Like their atomic counterpart in chemistry, SMs can exhibit periodic vibrations of their constitutive parameters \cite{Tchofo19, Zavyalov2009} (e.g. oscillations of the inter-soliton temporal spacing, of the optical phase difference, etc.). These SM systems have attracted much attention the last decade because they reveal the subtle dynamics at play in fiber lasers, and therefore call for more fundamental investigations with also possible applications in communication, and in all optical data processing and storage \cite{Pang2016,Liu2023}. This will be of course only possible if they can be controlled to an acceptable extent \cite{Liu2023}. From a fundamental prospect, SMs are a platform of choice to study the nonlinear dynamics of waves and their interaction. The existence of the SMs vibration is often ascribed to the formation of a limit-cycle attractor inside the laser cavity, which results in self-sustained oscillations even in the absence of any (operator-controlled) forced external excitation of the nonlinear oscillator \cite{Zavyalov2009}. In a sense, SMs are a very good example of nonlinear oscillators driven by the competition - or the balance- between dissipation and energy gain mechanisms. But fiber ring cavity lasers are extremely complex objects whose behaviors are set by multiple parameters \cite{ Zavyalov2009a, kutz98, Xian2022, Soto-Crespo2003, Jang2013}. If large efforts have been dedicated to both experimental measurements of the laser’s photonic state \cite{Goda2013,Herink2017} and to the control of the ring laser properties \cite{Nimmesgern21, Jang2013, He2019}, many unknowns still remain. Therefore quantitative controls and prediction strategies rely most of the time on machine-learning \cite{Girardot2020, Woodward2016}. This allows certain degree of control at the cost of intelligibility \cite{Haluszczynski2021}. A direct consequence is that most advanced optimized control and stabilization techniques still cannot be implemented \cite{Berkovitz2012, BoscainAnIT,Sugny2007}. Another strong limitation is related to the intrinsic stochastic nature of the laser cavity. Because the observed dynamics is actually the result of the balance between the various excitation mechanisms and the dissipative ones, it is alas not possible to measure these two effects separately. This in turn creates a halo of uncertainty when comparing theoretical modeling against experiment. Such a limitation is actually found in any system whose evolution laws are not known, and that are driven by noise (namely quasi random events).

\par We demonstrate in this article that it is possible to create, based on experimental data, a Langevin model \cite{Zwanzig1973, Darve2009, Brueckner2020} for the vibration of the inter-soliton distance.  Namely we derive a parsimonious set of equations \cite{Brunton2016, Floryan2022, Chen2022} that reproduces the deterministic (i.e. latent) dynamics of the SM, which we then complement by a noise model that encompasses the stochastic variability observed in the SM evolution. We show in particular that the reconstructed Langevin model allows the subsequent independent study of the dissipation and driving mechanisms, and of the noise properties. \pico{Previous studies involving numerical fitting and other deep learning techniques, particularly in the field of fiber lasers, usually considered the presence of noise as purely detrimental; and that it contains no valuable information.} If noise is indeed problematic for the data-driven construction of mathematical models because of the numerical artifacts it could create \cite{Prokop2024}; it however does not influence much the final result if the numerical aspects are appropriately taken care of \cite{Ermolaev2022}. We show here that noise can actually improves the quality and the robustness of the model. For this first proof of principle, we apply our strategy to the vibration pattern of a 2-soliton molecule in an ultrafast fiber ring cavity laser. We show how the Langevin model can be constructed; and the information it provides. \pico{As complex nonlinear oscillators where noise plays an important role, SMs constitute an interesting platform to test our method. The conclusions are, however, general and the technique presented here can be readily applied to any other dynamical systems.}

\section{Experimental dataset and SINDy technics}

\begin{figure}[htbp]
	\centering
	\includegraphics[width=13cm]{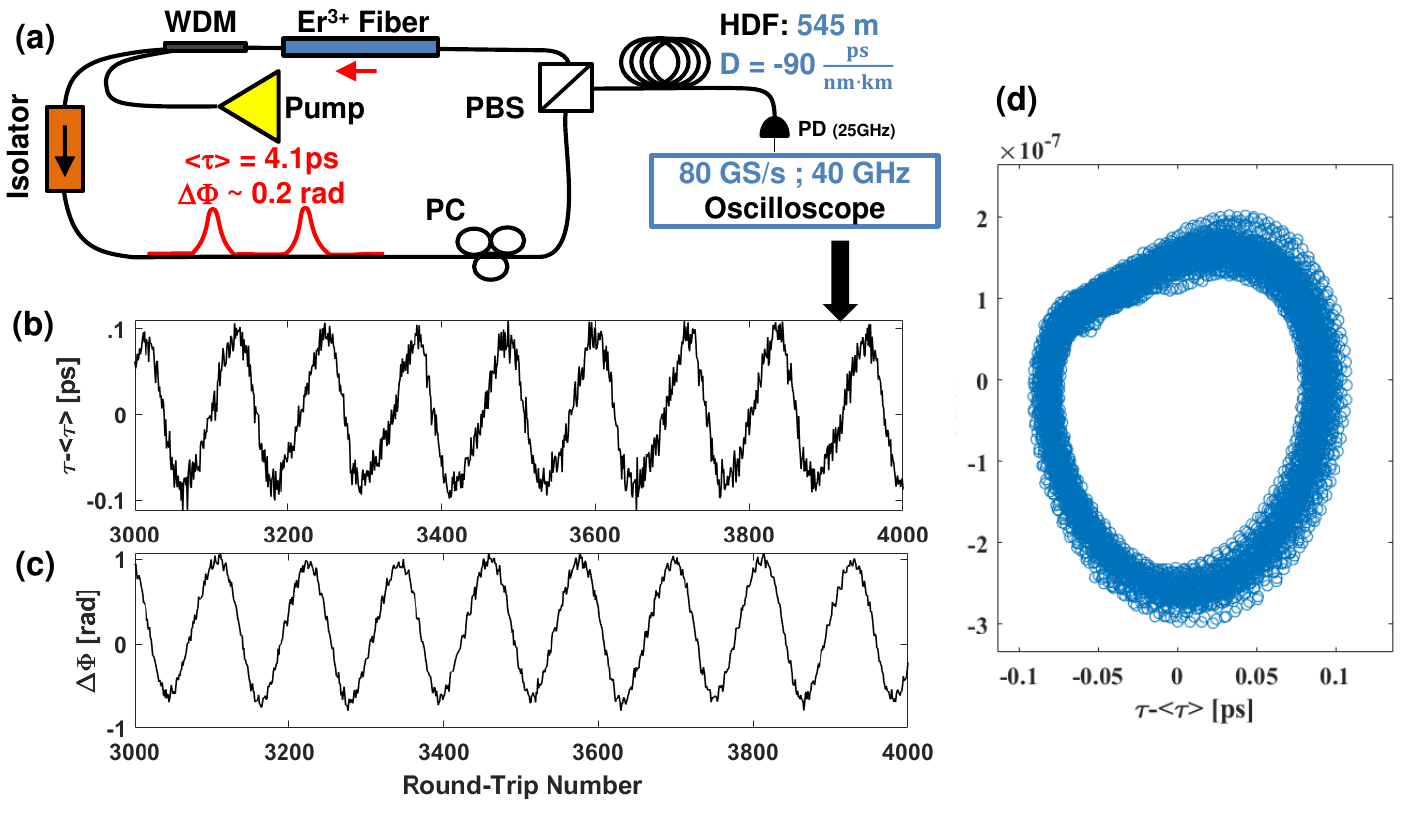}
	\caption{ (a) Experimental setup. PBS: Polarization Beam Splitter. PC: Polarization Controller. PD: Photodiode. HDF: Highly Dispersive Fiber (total dispersion of 50ps/nm). (b) Evolution of time separation $\tau$ between the two dissipative solitons. Its exhibits a characteristic oscillation with a period $T_0 = 117.6 ~ \text{RTs}$. $<\tau> = 3.48 ~ \text{ps}$. (c)  Evolution of the optical phase difference $\Delta\Phi$ between the two dissipative solitons. (d) Phase portrait in the plane $\{\tau,d{\tau}\}$ corresponding to (b).}
	\label{Fig:1}
\end{figure}


The experimental platform we consider in \fig{Fig:1}-(a) is an ultrafast fiber ring cavity laser composed of a 5-m long fiber ring laser cavity comprising 3 meters of Erbium doped fiber. The mode-locking mechanism comes from nonlinear polarization rotation in the single mode fiber (SMF) and a subsequent filtering by a polarizing beam splitter \cite{Hamdi2018}. This laser emits pulses at about 44~MHz repetition rate, and also can support SMs for higher pump settings. The SMs dynamics is recorded thanks to the Dispersive Fourier Transform technique (DFT) \cite{Goda2013,Godin2022,Herink2017}. For this, the laser output is dispersed in a -50~ps/nm Highly Dispersive Fiber (HDF), also known in the telecom industry as Dispersive Compensating Fiber - DCF. Finally, the DFT spectra are recorded by an 80~GS/s, 40~GHz electrical bandwidth, 8-bit depth oscilloscope. The photodiode has a 25~GHz bandwidth, and acts as a low pass filter for higher frequencies. The vibration pattern of the SM is shown in \fig{Fig:1}-(b-c): the parameters of interest are the time delay between the two pulses $\tau$ and their optical phase difference $\Delta\Phi$. Their evolution is observed over about 8,500 round-trips, as limited by the oscilloscope memory. It exhibits a characteristic periodic oscillation that presents however some minor fluctuations \cite{Hamdi2018, Herink2017, Tchofo19}. They are more visible on the phase portrait in \fig{Fig:1}-(d) which reveals indeed that the SM does not strictly stick to its limit cycle, but experiences cycle to cycle fluctuations. \pico{This results in a thick donut-shaped ring, instead of a well-defined loop}. We chose here to use the $\{\tau,d{\tau}\}$ representation to point out the applicability of our method to other systems, instead of traditional representation $\{\tau,\phi\}$ used for SMs. Indeed every other system may not allow the direct measure of two conjugated variables. In a more general context, the base of representation that is chosen may have a direct impact on the complexity of the subsequent mathematical model that is derived from the data, \pico{as discussed in more details in the next section.} 

\section{The SINDy model}

\begin{figure}[htb]
	\centering
	\includegraphics[width=13cm]{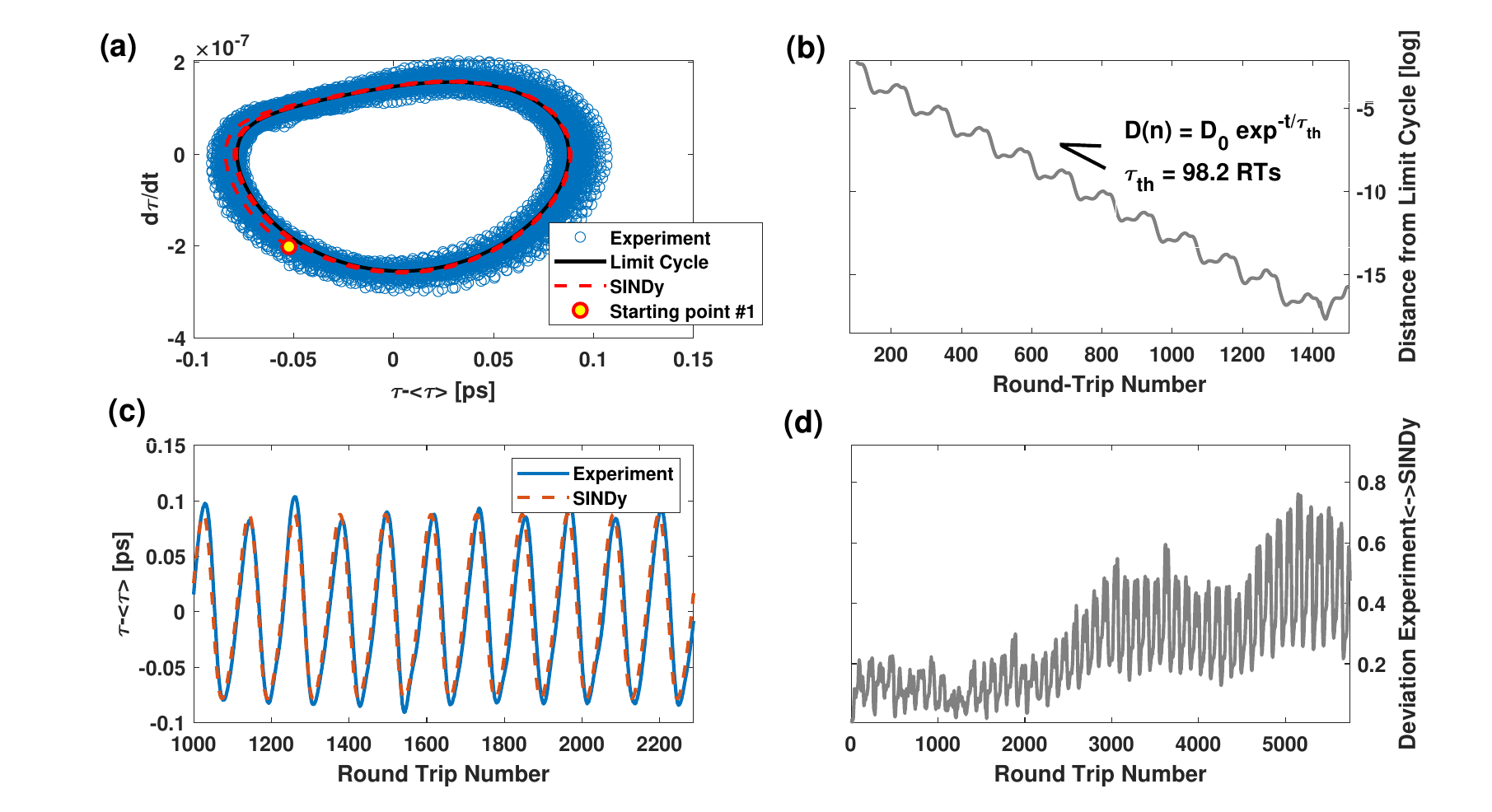}
	\caption{ (a)  $<\tau> = 3.48 ~ \text{ps}$. Blue-circles: Phase portrait of the SM oscillation in the $\{\tau; \dot{\tau}\}$ plane ($d{\tau}/dt = \dot{\tau}$). Black line: limit-cycle defined by \eq{eq2}. Red-dashed line: recovery dynamics for a noiseless model when the SM has been initially kicked from the limit-cycle, to the yellow dot, which serves as starting point. (b) Details of the recovery of the red-dashed trajectory presented in (a). The geometrical distance from the limit cycle evolves as an exponential decaying law with a characteristic thermalization time $\tau_{th}=98.2 \text{RTs}$. (c) Comparison between the experimental oscillations (solid blue) and the result of \eq{eq2} (dashed red). \pico{(d) Evolution of the deviation between the SINDy prediction and the actual experimental evolution. The deviation is measured as the euclidean distance in the phase portrait}.}
	\label{Fig:2}
\end{figure}

Following the Langevin’s formalism, the objective here is to model the experimental data using the following equation:
\begin{equation}
	\frac{d}{dt}\left[\begin{matrix} \tau \\{\dot{\tau}} \\ \end{matrix}\right] = 
	\left[\begin{matrix} \dot{\tau}\\ F(\tau , \dot{\tau})  \end{matrix}\right] + \left[\begin{matrix} 0 \\ \sigma_L(\tau , \dot{\tau}, t)  \end{matrix}\right]
	\label{eq:Langevin}
\end{equation}

Where $\dot{\tau}=d\tau/dt$, and $\sigma_L$ is the generalized Langevin’s force which is a null-centered random force. Note that its intensity could however depend on time and location $\{\tau , \dot{\tau}\}$ in the phase portrait. $F(\tau , \dot{\tau})$ is a nonlinear function that describes the deterministic evolution of the oscillator. It can be well approximated using the Sparse Identification of Nonlinear Dynamic (SINDy) algorithm \cite{Brunton2016}. The latter consists in finding the minimal set of elementary functions $f_i$ whose combination would approximate F to the desired accuracy. It corresponds to finding the most compact vector $\Xi=[\xi_1, ... ,\xi_p]$ that satisfies:

\begin{equation}
	\frac{d}{dt}\left[\begin{matrix} {\dot{\tau}}_1\\ \ \vdots \\ \ {\dot{\tau}}_N \\ \end{matrix}\right] = \left[\begin{matrix}f_1(\tau_1, {\dot{\tau}}_1) &\hdots& f_p(\tau_1, {\dot{\tau}}_1) \\ \vdots&&\vdots\\ f_1(\tau_N, {\dot{\tau}}_N) &\hdots& f_p(\tau_N, {\dot{\tau}}_N) \\\end{matrix}\right] \Xi
	\label{eq1}
\end{equation}

\begin{figure}[htbp]
	\centering
	\includegraphics[width=11cm]{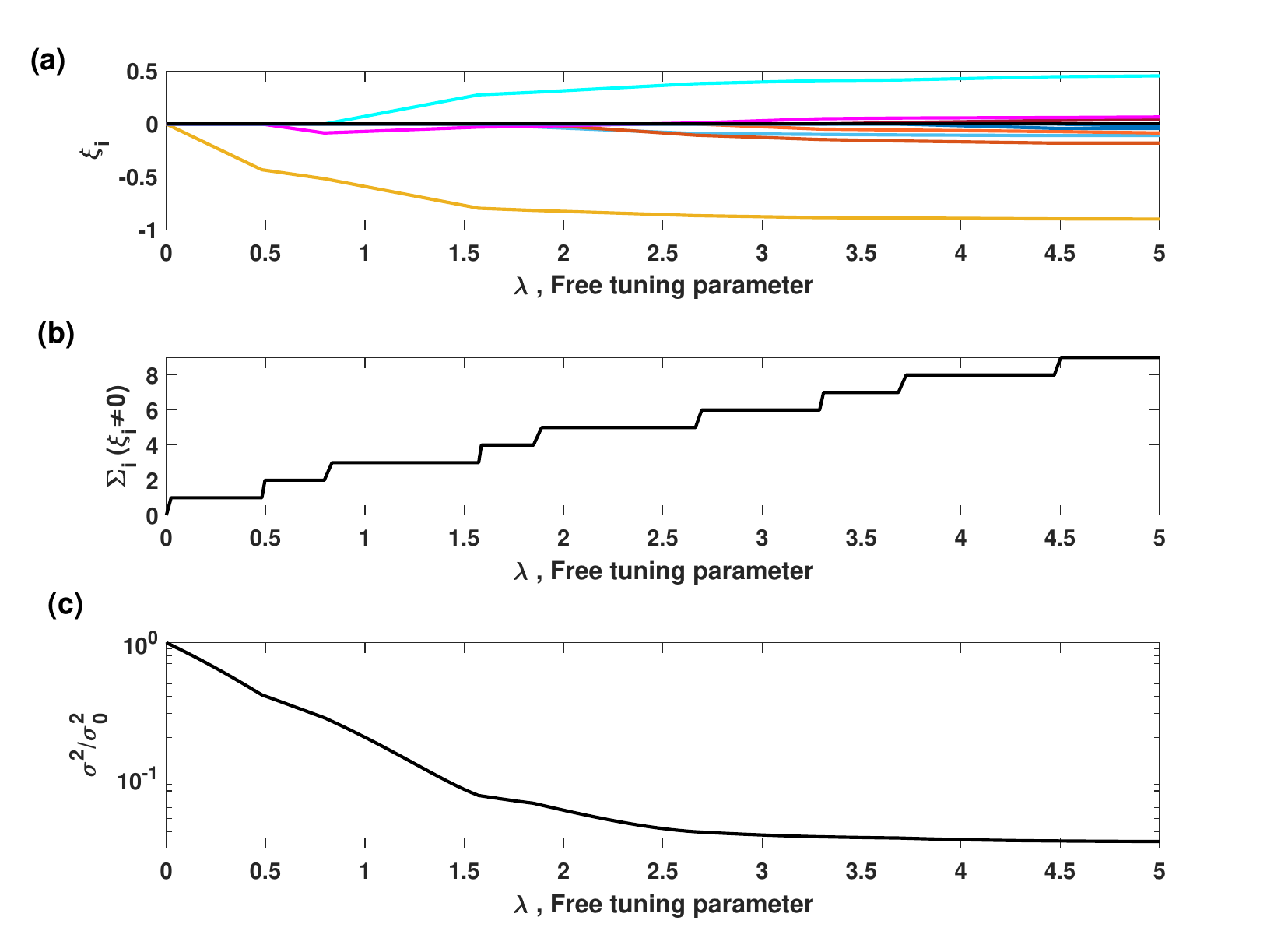}
	\caption{ \pico{(a) Evolution of the $\xi_i$ coefficients as function of $\lambda$, the free parameter controlling the LARS regression. (b) Number of non-zero coefficients. (c) Corresponding reconstruction error. $\lambda=3.7$ is chosen for the present work, corresponding to 7 non-zero parameters.}}
	\label{Fig:SINDY}
\end{figure}

The index notation $\tau_j$ represents the time records of $\tau(t_j)$. We define the normalized parameter $u = (\tau-<\tau>)/\sigma_{\tau}$, where $<\tau>= 3.48 \text{ps}$ and $\sigma_\tau = 0.060 \text{ps}$. We chose the library of linear functions $f_i$ to be composed of 36 polynomial functions of $u$ and $\dot{u}$ that span all the possible combinations, up to the order 7. This choice is arbitrary; and its quality could be improved by relying on a physics informed strategy where specific functions, which are inferred by theory to be potential candidate of the SM evolution, could be also included . \pico{This potential impact of arbitrary user-choices on the final result is completely analogue to the initial choice of the architecture for neural-networks. It is up to the user to be sure that the function to be approximated is indeed well encompassed by the system used for the training.} In this prospect, our approach to use only polynomial functions could be understood as performing a Taylor expansion of the actual SM's equation of evolution. We paid attention that the typical terms that one could expect to find in a simple anharmonic oscillator subject to friction loss (in particular the terms $u^3$ and $\dot{u}$) were indeed included in the library of test-functions. They were however not selected by the regression model. This indicates that the oscillator we are observing is actually of a different nature from the simple pendulum. That said, if a systems' evolution exhibits characteristics symmetry or scaling law, it is still a wise strategy to include in the library some functions that respect that specific symmetry or scaling law. We like to stress that the SINDy technique does not forcefully impose any specific model, hence avoiding any subsequent miss-interpretation of the system's dynamics. \pico{Besides, this technique involves the use of non-greedy algorithm with a $L_1$ penalty. This type of penalty is essential to result in parsimonious models because it tends to zero any irrelevant coefficients.} Regression is made here using the Least Angle Regression Algorithm (LARS \cite{Efron2004}), Other popular regression algorithms are the LASSO \cite{Tibshirani1996} and the Elastic-net \cite{Zou2006}, to cite a few. \pico{It is possible to evaluate the quality of the fit by checking which functions $f_i$ are selected and how their respective coefficients vary with the regression parameters. In the present case in \fig{Fig:SINDY}, the coefficients are varying smoothly with the tuning free parameter $\lambda$, and the functions with the highest polynomial order are also never selected. We see that for 7 non-zeros parameters}, the idealized dynamics of the SM can be reproduced with about \pico{3.5\%} accuracy. \pico{Considering that the SINDy algorithm is basically a least square error fitting, moderate zero-average random fluctuations have little impact on the final result, as it has been demonstrated \cite{Ermolaev2022}. That said,  an increased level of noise would greatly impact the precision of the regression coefficients of lesser magnitude. Performing the regression on isolated segments is essential to check that the SM does no undergo any notable change in its dynamics, as it would be characterized by large changes in the coefficients. For constant dynamics, the best strategy is nevertheless to perform the regression on the whole dataset so as to reduce the impact of noise to the minimum. The SM under study here can be modeled} using the following seven-parameter nonlinear equation: 

\begin{equation}
	\frac{d^2u}{dt^2} = \xi_1 u + \xi_2 u^2 + \xi_3 u^5  + {\dot{u}} (\xi_4 u + \xi_5 u^2) +  {\dot{u}}^2 (\xi_6 u^2 + \xi_7  u^3) 
	\label{eq2}
\end{equation}

$\Xi$ = (-0.758, -0.073, -0.01, 0.412, -0.171, -0.078, 0.011). Solution of \eq{eq2} is shown graphically in \fig{Fig:2}-(c). The equation indeed mimics well the experimental data, minus its local fluctuations. \pico{In particular the periodicity is well reproduced. Finer analysis of the divergence between the experimental evolution and the prediction from the SINDy model are presented in \fig{Fig:2}-(d). We see that the error remains about constant for about the first 2000 RTs, hence about 17 oscillations cycles; then it increases by steps. This indicates that the divergence is not caused by a drift, hence the incapacity of SINDy model to reproduce the correct periodicity, but can be ascribed to localized phase-shifts caused by the noise that offsets the experimental cycle. It takes about 8000RTs to reach about a complete cycle opposition.} It is interesting to note that the parameter $\xi_1$ in \eq{eq2} would correspond to a pure-sine linear oscillation of 112.9~RTs periodicity. Therefore we can conclude, based on the actual average $T_0=117.6~\text{RTs}$ periodicity which is observed over 72 full oscillations (about 8,500 Round-Trips), that the nonlinearity contributes notably to the oscillator's dynamics. It is also possible to derive from \eq{eq2} the shape of the limit cycle attractor (black line in \fig{Fig:2}-(a)). When compared to the actual experimental data, it clearly shows the impact of random fluctuations (noise) that forces the SM to explore a larger area around its limit-cycle. \pico{Therefore the fit is not strictly valid for reproducing the ideal oscillator dynamics (i.e. the limit cycle), but also retains some validity to describe the evolution of the SM when it is driven out of its idealized trajectory. The zone of confidence would correspond to the blue area in \fig{Fig:2}-(a) that is actually explored by the SM. Outside of this area the dynamics is inferred by analytical continuation, hence is becomes purely speculative for large deviations.} 

\section{SINDy: latent dynamics}

\par
\begin{figure}[htbp]
	\centering
	\includegraphics[width=13cm]{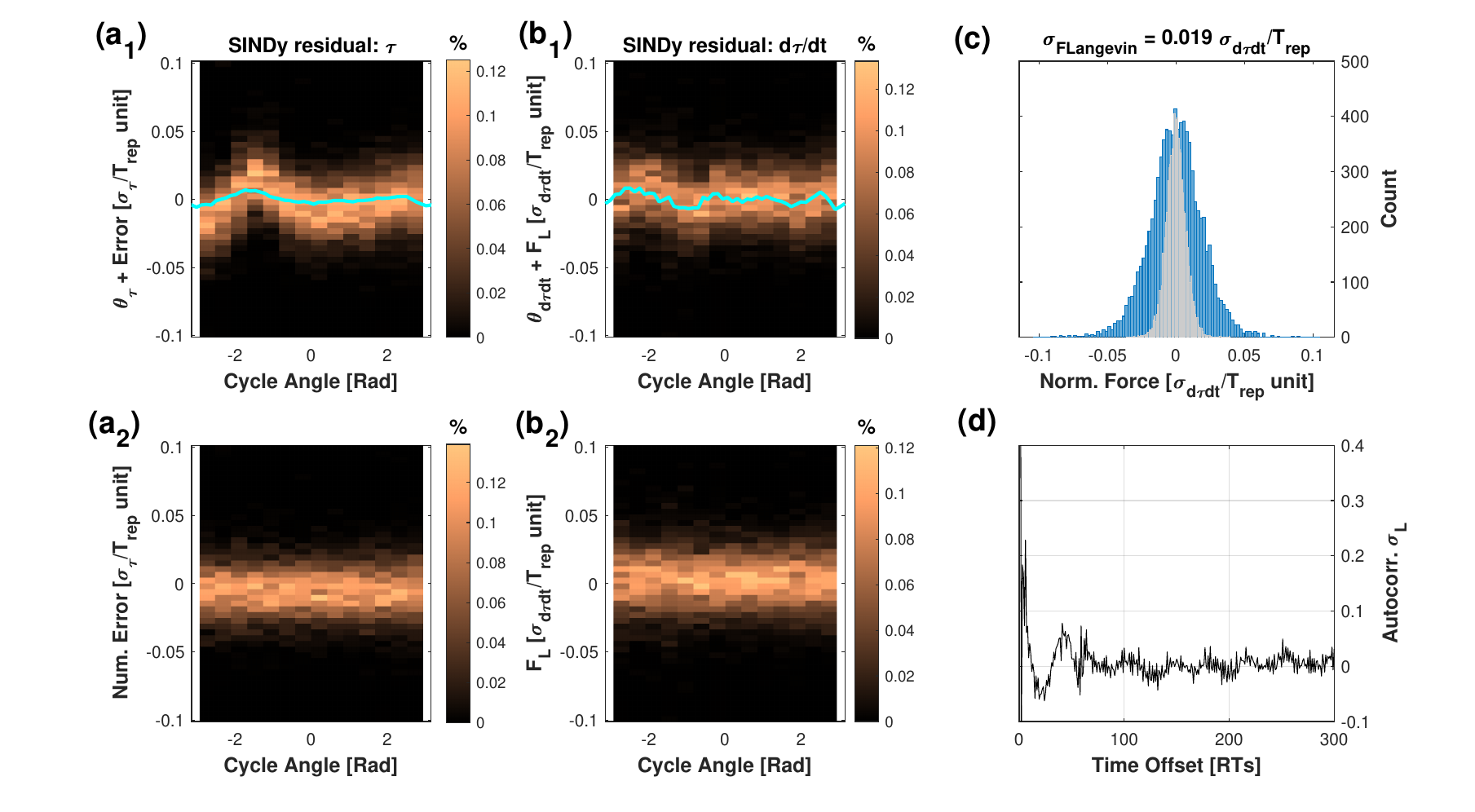}
	\caption{\pico{ ($a_1$,$b_1$) Results of \eq{eq:LangevinSINDY}. Cyan line indicates the average drift force, attributed to $\theta_{\tau}$ (resp. $\theta_{\dot{\tau}}$). ($a_2$-$b_2$) Resulting distribution of the Langevin's Force $\sigma_L$, according to \eq{eq:Langevin}. It is obtained after substraction of the drift $\theta$ in ($a_1$,$b_1$). By definition $F(\tau,\cdot{\tau}) =  F_{SINDy}(\tau,\cdot{\tau}) + \theta(\tau,\cdot{\tau})$. According to the Langevin's formalism, ($a_2$) must be null, and therefore is ascribed to numerical artifact. (c) Blue : Histogram of the Langevin's force shown in ($b_2$). Gray : histogram of the distribution of the numerical errors, construted from ($a_2$). (d) Correlation $<\sigma_L(t+t_0)\sigma_L(t)>_t/<\sigma_L(t)^2>_t $ as a function of the time offset $t_0$.}}
	\label{Fig:3}
\end{figure}

 A key point in the present analysis is that the noise forces the oscillator out of its equilibrium position. Therefore the oscillator can be observed over a larger area than its limit-cycle without having to actually \pico{drive forcefully the oscillator out of its nominal trajectory. SM are indeed fragile soliton complexes that could be easily destroyed. Aside from resulting to a fit with a larger zone of validity,} this subcritical approach has also other advantages. Each noise event can indeed be interpreted as a random kick that excites the system, causing it to deviate from its stable recurrent dynamics. In case the random kicks can be tracked individually, it is then possible to observe the systems' response to a perturbation without having to actually perturb it. As a result this approach is non-invasive, hence it does not risk destroying any fragile dynamics equilibrium that may exist\pico{, nor requires a mean to actually excite the system. That said because the SM is under constant perturbation from noise, it is not possible to observe the complete relaxation dynamics. This would require being able to switch off completely the noise for an extended time, which is not possible experimentally. But using the numerical model we just derived as a numerical twin, it is now possible to investigate how the SM would have relaxed in absence of noise from a position away from the limit-cycle (yellow circle in \fig{Fig:3}-(b))}: following the noiseless model defined in \eq{eq2}, the SM will then dissipate the excess of energy and spiral down back to its limit-cycle.

The (normalized) geometrical distance from the limit-cycle is shown in \fig{Fig:2}-(b). The SM recovers to its nominal steady dynamics with a relaxation time $\tau_{th} = 98.23~\text{RTs}$. \pico{This set the typical time-range of excitations which the SM would be able to respond to.} Note that this characteristic time is about equals to the actual oscillator periodicity (117 RTs). Faster excitations would leave no time for the oscillator to adapt to them; they could destroy the SM or cause its transfer to another limit-cycle (provided the latter exists) \cite{Kurtz2019}. Similarly, \eq{eq2} can be further analyzed to determine the loci of the driving and the dissipation mechanisms, the strength of the potential confining the SM, etc. Interestingly, for the present case, the sole equilibrium position would correspond to static molecule ($u=0 <=> \tau=<\tau>=C^{onst}$).

That said, the SINDy analysis still indicates that the SM does not spontaneously oscillate around its static equilibrium position: the sustainability of the current SM system is indeed due to the velocity contribution $\dot{u}$, which is always repulsive for the small excursion $|\tau-<\tau>|<0.14~\text{ps}$ experienced here by the SM. As a consequence, we can infer that the existence of the oscillation is actually intrinsic to the existence of an initial drift velocity between the two pulses during the formation of the molecule. An important level of noise can also prevent the molecule to remain static. This rises the question regarding the existence of specific classes of SMs depending on whether the oscillation occurs around a static equilibrium position, or if it is intrinsically linked to the stability of the SM. \pico{Answering such a question would require to perform the current analysis on numerous SM records at different but controlled laser’s parameters. Note that the physical parameters (e.g. type and parameters of the gain medium, etc.) influence the laser dynamics is a non-trivial manner; and their impact is not known yet quantitatively. This would then require more complex fiber laser involving fine stabilization; and exceed the scope of the present proof-of-principle demonstration}.

\section{Impact of noise on the dynamics: the Langevin model}

\par The SINDy analysis \cite{Brunton2016} is an important tool in order to infer latent (i.e. deterministic) dynamical features that cannot be measured directly because of the presence of noise. It can also be used to extract stochastic properties. 

According to \eq{eq:Langevin}, the Langevin force $\sigma_L$ \pico{causes the SM to experience a dynamics different from the latent dynamics, approximated by the SINDy regression, as discussed in the previous session. The real latent dynamics differs of course slightly from the SINDy ones, so that \eq{eq:Langevin} can be rewritten as:} 

\begin{equation}
	\frac{d}{dt}\left[\begin{matrix} \tau \\{\dot{\tau}} \\ \end{matrix}\right] = 
	\left[\begin{matrix} \dot{\tau} \\ F_{SINDy}(\tau , \dot{\tau})  \end{matrix}\right] + \left[\begin{matrix} \theta_{\tau}(\tau , \dot{\tau}) \\ \theta_{\dot{\tau}}(\tau , \dot{\tau})+ \sigma_L(\tau , \dot{\tau}, t)  \end{matrix}\right]
	\label{eq:LangevinSINDY0}
\end{equation}

\pico{Where $\theta(\tau , \dot{\tau})$ is the error between the SINDy regression and the real latent dynamics. Comparing for each round-trip the divergence of evolution $\{\delta\tau , \delta\dot{\tau}\}$ between the actual SM's evolution and the SINDy projection results for small divergences into the following linearized equation:} 

\begin{equation}
	\frac{d}{dt}\left[\begin{matrix} \delta\tau \\ \delta{\dot{\tau}} \\ \end{matrix}\right] = 
	\left[\begin{matrix} 1 & 0 \\ \partial_{\tau}F_{SINDy}  & \partial_{\dot{\tau}} F_{SINDy} \end{matrix}\right] \left[\begin{matrix} \delta\tau \\ \delta{\dot{\tau}} \end{matrix}\right] + \left[\begin{matrix} \theta_{\tau}(\tau , \dot{\tau}) \\ \theta_{\dot{\tau}}(\tau , \dot{\tau})+ \sigma_L(\tau , \dot{\tau}, t)  \end{matrix}\right]
	\label{eq:LangevinSINDY}
\end{equation}

\pico{With the observation of the divergence between the experimental data and the SINDy model, \eq{eq:LangevinSINDY} can be solved for $\theta$ and $\sigma_L$. The results are presented graphically in \fig{Fig:3}-$(a_1,b_1)$, where the histograms of $\theta + \sigma_L$ are presented depending on the angular position along the limit cycle (trigonometric orientation, zero angle at East direction). By definition, the Langevin's force is random, and therefore the average of the histogram, hence the average drift, can be attributed to $\theta$ (pink line in \fig{Fig:3}-$(a_1,b_1)$). The resulting $\sigma_L$ is presented in \fig{Fig:3}-$(b_2)$: its amplitude and distribution does not depend on the loci. Moreover it exhibits a Gaussian distribution (\fig{Fig:3}-(c)), uniform over the whole phase portrait, hence $\sigma_L(\tau, \dot{\tau},t) = \sigma_L(t)$. In the canonical Langevin model, the remaining distribution \fig{Fig:3}-$(a_2)$ would be zero. It serves here as an estimate of the numerical artifacts caused by the noise and the subsequent numerical processing. In particular the computation of second order time derivative on noisy signals is known to be a source of problems. These numerical artifacts are however by three times weaker than the Langevin perturbation, so we can conclude that both our numerical treatment and the presented analysis is legitimate.} The magnitude of the Langevin's force can be normalized to the dissipation capability of the limit-cycle. Its lesser magnitude, combined with the absence of extreme rogue events, indicates that the current SM is operated in a stable regime, away from any bifurcation point.

\section{Discussion}

\par
\begin{figure}[htbp]
	\centering
	\includegraphics[width=13cm]{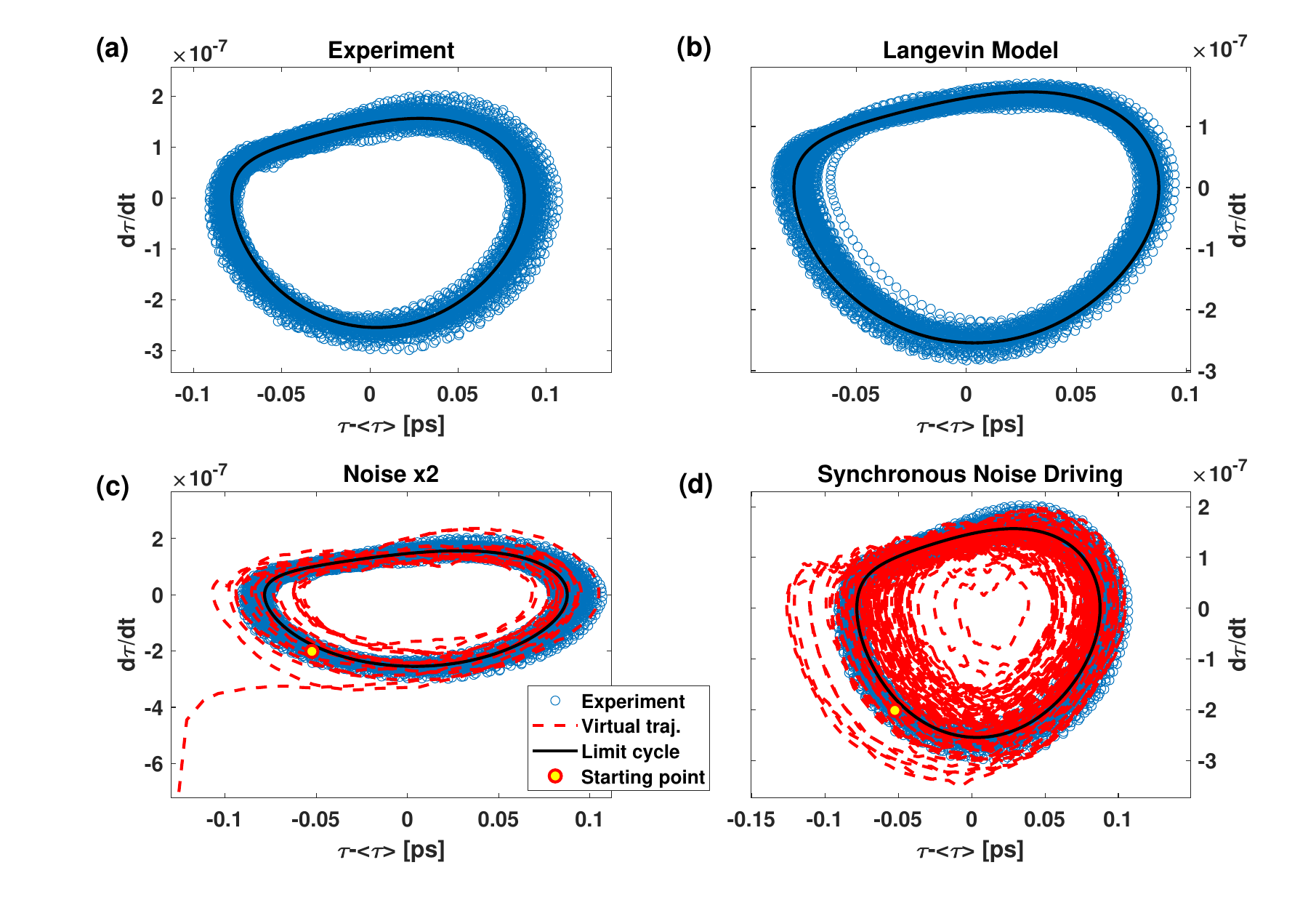}
	\caption{\pico{ (a) Observed dynamics.(b) Nominal Langevin Model as defined by \eq{eq:LangevinSINDY0}. (c) red-dash : Result of Langevin simulation where the noise amplitude (i.e. $std(\sigma_L)$) has been multiplied by two. The SM breaks appart quickly. (d) red-dash : Result of Langevin simulation where the frequency of the noise correlation has been set to the oscillation period. The SM breaks appart eventally, after a rather long evolution.}}
	\label{Fig:4}
\end{figure}

 We would like to stress that the method we propose here never assumes any peculiar shapes regarding the noise distribution. This is a very practical advantage over other stochastic optimization methods like the Fokker-Planck optimization \cite{Risken1996, Cheng2006}. However the naive approach that we use here requires to track the evolution of individual particles, which could be problematic for other systems composed of numerous indiscernible particles. The possibility to characterize experimentally the strength and distribution of fluctuations would allow to get the SM progressively arbitrarily close (or far away, depending on the application) from any transition. Further information can be obtained from the time correlation of the noise, as shown in \fig{Fig:3}-(d).  In details, we first see at small evolution time a strong positive correlation followed by a strong negative anti-correlation. On longer timescale, the noise retains about 3\% correlation/anti-correlation oscillations with about 1.1 MHz periodicity. We ascribe this behavior to the intrinsic dynamic of the gain medium where the evolution of the gain level is inversely correlated to the variation of the optical power circulating within the cavity. \pico{Any power drop in the cavity would result into less power taken at each round-trip from the gain medium; in turn the latter can build up extra energy and gain, which results later into an increase of the optical intra-cavity power. Same reasoning works in case of a transient increase of optical power.} Negative delayed feedback is known to create oscillations, with typical ${\mu}s$-timescale for ion-doped fibers \cite{Desurvire1989, Kuroda2015, Khamis2019}. For cavity laser with active control, this residual oscillation could eventually be decreased further if the right control strategy is employed; and any instability created by the feedback control itself could also be easily identified by this manner. This demonstrates further the potential of our method of analysis.\\Finally we present in \fig{Fig:4}-(b) the result of the full Langevin model \eq{eq:Langevin} that combines the latent dynamics and the noise model as defined in \fig{Fig:3}-(c,d). This can be directly compared with the full experimental data shown in the left panel \fig{Fig:4}-(a). We see that the main features are reproduced correctly. The differences come from the over-simplification of the oscillator model through the SINDy regression, where parts of the confinement potential cannot be approximated analytically with enough accuracy. The asymmetry of the confining potential is directly reflected on the the skewness of the SM probability of occupation around its limit cycle. This problem can be corrected through the further addition of extra small non-analytical corrective terms into $\theta(\tau, \dot{\tau})$  in \eq{eq2}. This is subject to the fact that there are enough sampled points at each location to determine the latter with precision. This increase in numerical precision would however come at the cost of the pure analytical description as provided by the SINDy algorithm. This is actually an inherent weakness of the SINDy method where good knowledge of the system is required in order to select the right set of functions for the optimization step. \pico{That said the numerical twin which we just created brings also a strong advantage in the sense that it is now possible to experience virtually different configurations and interplay scenarii between the noise and the latent dynamics, as presented in \fig{Fig:4}-(c,d). Firstly, increasing the amplitude of the Langevin force, hence noise, by a factor two would result in the destabilization of the SM which would break apart 80\% of the time within 8500 RTs (\fig{Fig:4}-(c)). No breaking is observed at nominal noise level. This indicates that while the SM is stable, a slight increase of the noise would have dramatic impact. We then investigated in \fig{Fig:4}-(d) the impact of the frequency of the residual noise temporal correlation. At the nominal noise and correlation levels, but with a correlation frequency synchronized with the SM oscillation period, the SM is gradually exploring a larger and larger area. Eventually the SM breaks also apart, but after a longer evolution than in the case of an increased noise intensity. We can conclude from this that any remaining noise correlation, albeit weak, seems to prevent the existence of a SM that would be synchronized to it.}

\section{Conclusion}

\par To conclude, we constructed a Langevin model for the accurate description of the vibration pattern of a soliton molecule (SM). Interestingly, the presence of fluctuations does not hinder the regression process, but on the contrary makes it stronger by exploring part of the parameters space outside the limit-cycle of the nonlinear oscillator. The key advantage of our approach is that, in a subsquent analysis after the mathematical model is constructed, it is possible to differentiate between the latent evolution of the nonlinear oscillator and its stochastic properties. Therefore, characteristic features such as the strength of noise or the intrinsic oscillator’s relaxation dynamics can be inferred, albeit they cannot be measured independently. Moreover the SM properties are inferred in a sub-critical approach that does not require to actively perturb or probe the laser cavity. \pico{But thanks to the numerical twin that is constructed, it is still possible to explore further the impact of noise, including the onset at which it could alter notably the SM dynamics}. These general features are intrinsically linked to the global fitting procedure that we propose, which encompasses both features of the latent dynamics and of the noise. Outside the field of fiber lasers, we believe that the construction of a Langevin model, based on a SINDy regression, could be applied to any system composed of trackable particles. More specific to systems for which the final objective is to control them, performing a parametric Langevin analysis against a few of the parameters that can be controlled would allow a deterministic control. Relying on subsequent optimal control techniques would be a competitive approach to the use of deep-learning techniques. The use of the latter is trendy for the control of black-box systems, but their training requires large data set which is not always possible to obtain. Moreover the reliability of the resulting control cannot be assessed; and any deviation from the system's points of operation tested during the training could lead to failure. Here the langevin's analysis provides estimate for the stability zones; and ensures the absence of uncontrolled transition states while changing dynamically point of operation. We believe that it would allow better control and comprehension of complex laser systems.\\ For this first demonstration, we applied our analysis on the $\left\{\tau, \dot{\tau}\right\}$  parameter space to illustrate the generality of our method to any situation where a time signal is recorded. Note that in specific situations where conjugated parameters could be directly recorded, it would then remove the necessity to compute a numerical time derivative, which is always a problematic matter in presence of noise, hence improve the overall quality of the analysis.


	\section*{Funding}
	This work was supported by the French ANR funded CoMuSim  (ANR-20-CE24-0007) and OPTIMAL projects (ANR-20-CE30-0004).
	
	\section*{Acknowledgments}
	P.C. thanks Omri Gat for fruitful discussion.
	
	\section*{Disclosures}
	The authors declare no conflicts of interest.
	
	\section*{Data Availability Statement}
		Data underlying the results presented in this paper may be obtained from the authors upon reasonable request.
	
	\section*{Contribution}
	A.C and P.C acquired the experimental data. A.S and P.C. performed the numerical analysis. All authors contributed to the writing of the manuscript. This project has been supervised by P.C..


\bibliographystyle{elsarticle-harv} 
\bibliography{Biblio_PiCo.bib}





\end{document}